\pdfoutput=1
\documentclass[useAMS,usenatbib]{mn2e}
\usepackage{amsmath, float}
\usepackage{txfonts}
\usepackage{fleqn}   
\usepackage{graphicx}

\newcommand{\be}{\begin{equation}}
\newcommand{\ee}{\end{equation}}
\newcommand{\bdm}{\begin{displaymath}}
\newcommand{\edm}{\end{displaymath}}

\newcommand{\bea}{\begin{eqnarray}}
\newcommand{\eea}{\end{eqnarray}}
\newcommand{\ba}{\begin{align}}
\newcommand{\ea}{\end{align}}

\title[On the MHD modeling of the Crab Nebula radio emission]
{On the MHD modeling of the Crab Nebula radio emission}
\author[B. Olmi, L. Del Zanna, E. Amato, R. Bandiera, N. Bucciantini]{B. Olmi$^{1,2,3}$\thanks{E-mail:
barbara.olmi@unifi.it} , L. Del Zanna$^{1,2,3}$, E. Amato$^{2}$, R. Bandiera$^{2}$, N. Bucciantini$^{2,3}$\\
$^{1}$Dipartimento di Fisica e Astronomia, Universit\`a degli Studi di Firenze, Via G. Sansone 1, 50019 Sesto F.~no  (Firenze), Italy\\
$^{2}$INAF - Osservatorio Astrofisico di Arcetri, Largo E. Fermi 5, 50125 Firenze, Italy\\
$^{3}$INFN - Sezione di Firenze, Via G. Sansone 1, 50019 Sesto F.~no  (Firenze), Italy}

\begin{document}
 
\date{Accepted/Received}

\maketitle

\label{firstpage}

\begin{abstract}
In recent years, it has become a well-established paradigm that many aspects of the physics of 
Pulsar Wind Nebulae (PWNe) can be fully accounted for within a relativistic MHD description. 
Numerical simulations have proven extremely successful in reproducing the X-ray morphology 
of the Crab Nebula, down to very fine detail. Radio emission, instead, is currently one of the most obscure 
aspects of the physics of these objects, and one that holds important information about pulsar properties 
and their role as antimatter factories. Here we address the question of radio emission morphology 
and integrated spectrum from the Crab Nebula, by using for the first time an axisymmetric
dynamical model with parameters chosen to best reproduce its X-ray morphology. 
Based on our findings we discuss constraints on the origin of the radio emitting particles.
\end{abstract}


\section{Introduction}

Pulsar Wind Nebulae arise from the interaction of the relativistic wind emanating from a rapidly 
rotating neutron star with its surroundings. Effective confinement of the pulsar wind by the debris 
of the parent supernova explosion induces the formation of a termination shock at which the wind 
is slowed down and its energy is transformed into that of the magnetized relativistic plasma 
responsible for the nebular emission. The latter is mostly non-thermal and very broad-band: 
the class prototype, the Crab Nebula, was one of the first synchrotron sources ever identified 
and shows emission over more than 15 decades in frequency \citep[e.g.][]{Hester:2008}. 

PWNe collect most of the rotational energy lost by the parent pulsars. As such they are a 
privileged location to look for answers to old and new questions in pulsar physics. 
While pulsars are thought to be the primary leptonic antimatter factories in the Galaxy, 
a big open question concerns the exact amount of pair production in their magnetospheres, 
the so-called pair multiplicity. In this time of new observational and theoretical developments 
on the subject, modeling of PWNe is likely to provide the most solid constraints \citep{Bucciantini:2011}.
   
In the last few years the axisymmetric relativistic MHD models of PWNe 
\citep{Komissarov:2004a, Del-Zanna:2004, Del-Zanna:2006} have proven extremely successful at accounting 
for the properties of these objects as observed in the optical/X-ray band. 
These models identify in the anisotropy of the pulsar outflow, more energetic in the equatorial plane 
of the pulsar rotation than along the polar axis, the origin of the jet-torus morphology revealed 
by \textit{Chandra} in many of these nebulae \citep{Weisskopf:2000,Gaensler:2006}. In addition they 
provide a simple explanation for the time variability observed in the inner region of the 
Crab Nebula \citep{Volpi:2008, Camus:2009} and for much of the finer scale structure of the emission.

So far, MHD modeling of PWN radiation has focused on high energy photons, 
while completely ignoring the low frequencies, in spite of the fact that the correct interpretation 
of PWN radio emission is of fundamental importance for understanding many aspects 
of PWN and pulsar physics. The big question to assess is whether the radio emitting particles 
are continuously injected in the nebula as part of the pulsar wind and accelerated at 
the termination shock, or they can rather have a different origin and acceleration site. 
Indeed, if the radio emitting particles are continuously injected in the nebula together with 
the optical/X-ray emitting ones, then the inferred pulsar multiplicities are typically larger 
than current pulsar theories are able to explain \citep{Hibschman:2001}. For example, in the case of the Crab Nebula, 
including low energy particles in the pulsar outflow one finds the pulsar multiplicity 
$\kappa \!\sim\! 10^6$ and a wind Lorentz factor, averaged over solid angle, 
$\gamma\!\sim\!10^4$ \citep{Bucciantini:2011}. If radio emission comes from particles of a different origin, 
then, from higher frequency observations one deduces $\gamma\!\sim\!10^6$ and 
$\kappa \!\sim\! 10^4$ \citep[e.g.][]{Kennel:1984,Gaensler:2002}. The value of these two parameters has 
also important consequences in terms of constraining the acceleration process at 
work at the wind termination shock \citep{Sironi:2011,Arons:2012}.

The radio emitting particles, with their large number and long life-times against synchrotron losses, 
could also be relics of earlier times, when the pulsar wind was different from now \citep{Atoyan:1996}. 
Or they could come from somewhere else than the pulsar wind, with a possible source being 
the thermal filaments penetrating the relativistic bubble \citep[e.g.][]{Komissarov:2013}. In the latter case 
the electrons would have to be accelerated locally to relativistic energies. A possible mechanism to this
purpose relies on scattering off local turbulence (Fermi II type process), providing, at the same time, 
acceleration and spatial diffusion, so as to guarantee a smooth distribution of the radio emission
throughout the nebula, despite the presence of enhanced emission associated with the filaments.
In the presence of efficient scattering by local turbulence, a rather uniform distribution of particles
would also be expected in the case in which the particles are re-accelerated relics, namely they were
injected at the termination shock only for a short time during the PWN lifetime, but then re-energized
by the interaction with turbulence.

In this article we investigate the problem of the origin of radio emitting particles. 
For the first time we compute radio emission maps based on the flow structure that arises from the 
2D axisymmetric MHD simulation that best reproduces the Crab Nebula X-ray emission. 
We compare the emission morphology under three different hypotheses: A) assuming that 
radio particles have always been accelerated at the termination shock; B) taking their distribution as uniform 
in the nebula, as could be possibly expected if after being injected, at the shock or by the filaments, 
they were (re-)accelerated by ambient turbulence; C) assuming that they were only injected at the termination 
shock in the nebula for a relatively short time after the supernova explosion and with no further re-acceleration.

One piece of information that seems to point towards a common origin of the radio emitting and 
higher energy particles is the observed continuity of the integrated emission spectrum of the Crab Nebula, 
from which continuity of the spectrum of the emitting particles has traditionally been inferred. 
For other PWNe, the data are just not good enough to allow to assess this issue, also due to 
the scarcity of observations of these objects in the IR and optical band, which is where the 
transition between different spectral slopes usually occurs. 
In the presence of a uniform magnetic field strength, as assumed by 1-zone models, 
a smoothly joined emission spectrum between radio and optical implies a smooth spectral 
distribution of the emitting particles. In the presence of a spatially inhomogeneous field, 
this is not a straightforward inference,
since the volume occupied by particles of different energies is different (larger at lower energies) 
and so is in principle the field strength. We investigate the paradigm of continuity 
and smoothness of the particle spectrum in the Crab Nebula and show that the IR bump 
of thermal emission associated to dust could easily hide a possible gap between radio 
and optical emitting particles.

The paper is organized as follows: in Sec.~\ref{sec:setup} we describe the adopted pulsar 
wind model, the simulation set-up, the assumptions on the particle spectrum, and our procedure 
to calculate the emission; in Sec.~\ref{sec:res} we present our results and discuss the implications 
of different models, and we conclude in Sec.~\ref{sec:concl}.

\section{Pulsar wind model and particle emission}
\label{sec:setup}

The initialization and set up of our 2D numerical simulations is very similar to that described 
in our earlier papers: here we double the resolution and relax the assumption of symmetry at the equator. 
The physical domain ranges from $r_\mathrm{min}=0.05$ ly to $r_\mathrm{max}=10$~ly, 
with logarithmic stretching ($dr\propto r$). 
The angular domain ranges from 0 to $\pi$, with reflection conditions at the polar axis. 
A spatial grid with 800 cells in the radial direction ($\simeq 350$ points per decade) 
and 400 cells in the polar angle $\theta$ is used. We take 
$\varv_\phi = B_r = B_\theta \equiv 0$ and the equation of state for an ideal gas with 
$\gamma=4/3$, appropriate for the ultra-relativistic limit.

The simulation box is initially divided into four different regions. 
From outside in we set up: the fully ionized ISM with $n=1~{\rm cm}^{-3}$; 
the cold supernova ejecta, with total mass corresponding to $M_\mathrm{ej}= 6 M_\odot$ 
and expansion profile $\varv\propto r$; a \emph{primordial} hot bubble
 that during the evolution will merge in the actual PWN (just used for numerical stability: 
 it largely avoids numerical diffusion at the contact discontnuity); the cold, magnetized pulsar wind 
expanding with a relativistic Lorentz factor $\gamma\equiv\gamma_0=100$ and continuously 
injected from the inner boundary at $r_\mathrm{min}$. 
This value of $\gamma_0$ is much lower than what
expected to be appropriate for the Crab pulsar wind \citep[e.g.][]{Kennel:1984}, but still 
high enough to guarantee that the flow is highly relativistic, in which case the post-shock 
dynamics is independent on the exact value of the Lorentz factor.

The structure of the pulsar wind is crucial in determining the shape of the PWN. 
We assume an energy flux that roughly depends on space as predicted by 
\textit{split monopole} models of the pulsar magnetosphere \citep{Michel:1973}, 
namely  $F(r,\theta) = c ( n m_\mathrm{e} c^2 \gamma_0^2 + B^2/4\pi ) \propto r^{-2} \sin^2\theta$, 
where $B\equiv B_\phi$ is the embedded toroidal magnetic field and $n$ is the 
number density of the outflow, but we remark that recent 3D models seem to indicate 
a different behaviuor for oblique rotators ($\sim \sin^4\theta$) \citep{Spitkovsky:2006a,Tchekhovskoy:2013}.
The precise spatial dependence of the energy flux is taken as
\begin{equation}
\label{eq:total_flux_2}
F(r,\theta)= \frac{L_0 }{4 \pi r^2} \mathcal{F}(\theta)\,  , 
\quad \mathcal{F}(\theta)=\frac{\alpha + (1-\alpha)\sin^2\theta}{1-(1-\alpha)/3}\,,
\end{equation}
where $L_0=5 \times 10^{38}$~ erg~s$^{-1}$ is the pulsar spin-down luminosity, which 
we take as constant in time, and $\alpha \ll 1$ is the 
\emph{anisotropy parameter}, which controls the ratio between polar and equatorial energy flux. 
The magnetic field is assumed as
\begin{equation}
\label{eq:magnetic-field-1}
B(r,\theta)= \sqrt{\frac{\sigma_0 L_0}{c}}\frac{\mathcal{G}(\theta)}{r}\, ,\quad
\mathcal{G}(\theta)=\sin\theta \tanh \left[ b\left(\frac{\pi}{2}-\theta \right)\right],
\end{equation}
where $\sigma_0$ defines the magnitude of $B$. The function $\mathcal{G}(\theta)$ 
is chosen having in mind the split-monopole dependence on latitude, but also the fact that the wind 
is expected to be striped \citep{Coroniti:1990} and dissipation is likely to take place between 
the stripes within a belt around the pulsar rotational equator. The parameter $b$ relates to the width 
of this belt: for large values of $b$ the pure split-monopole is recovered, while for $b\sim 1$ dissipation 
modulates the field strength at all latitudes \citep{Del-Zanna:2004}. The wind magnetization is defined 
as $\sigma= B^2/(4\pi nm_\mathrm{e} \gamma_0^2 c^2)=\sigma_0 \mathcal{G}^2/(\mathcal{F}-\sigma_0\mathcal{G}^2)$ 
and depends on $\theta$ alone. Finally, Eqs.~\ref{eq:total_flux_2} and \ref{eq:magnetic-field-1} lead to
\be
n(r,\theta) = \frac{L_0}{4\pi c^3 m_\mathrm{e} \gamma_0^2} \frac{1}{r^2}
\left[ \mathcal{F}(\theta) - \sigma_0 \mathcal{G}^2(\theta) \right].
\label{eq:n}
\ee

The synchrotron spectrum of the Crab Nebula can be described as produced by particles 
with a broken power-law distribution in energy \citep{Amato:2000}. In the following we assume that the particles 
responsible for the optical/X-ray emission are, in all cases, constantly accelerated at the shock, 
while the radio emitting particles can have different origins: be part of the pulsar outflow and 
constantly accelerated at the shock (Case {\it A});  uniformly distributed in the nebula (Case {\it B});
resulting from a burst of acceleration at the shock, only lasting for a limited time (of order 10\% of the nebular age), 
and then advected in the nebula, with no spatial diffusion or further re-acceleration (Case {\it C}). 

We would like to point out, right from the beginning, that Case {\it C} corresponds to a model of relic particles 
that were injected in the nebula at some early time, when the pulsar was spinning down in a different way, 
either in terms of overall $L_0$ or in terms of producing a different multiplicity. This pure
relic model, in which the particles are not re-accelerated in the nebula after their initial early injection, is only considered to the purpose of illustrating how the emission morphology would change under this extreme assumption.
Indeed, a model of relics that does not conflict with any energetic requirement is likely to give rise to 
a particle distribution close to that assumed in Case {\it B}. Here the underlying physical picture is that 
of particles being scattered and accelerated by turbulence in the nebular volume. No matter whether 
they are extracted from the thermal bath of the filaments or whether they were injected at the shock at early time, 
the interaction with turbulence that ensures (re)acceleration also smoothes out all the spatial gradients.

Since both model {\it A} and {\it B} give results in reasonable agreement with the data, in the following 
we discuss them in parallel, while treating Case {\it C} separately.

The general form of the distribution function of newly injected particles in the nebula is given as
\begin{equation}
\label{eq:f_0}
f_{0s}(\epsilon_0)= \left(p_s-1\right)\frac{K_s}{\epsilon_s^\mathrm{min}} 
\left(\frac{\epsilon_0}{\epsilon_s^\mathrm{min}}\right)^{-p_s}, \!\!
 \quad \epsilon_s^\mathrm{min} \le \epsilon_0 \le \epsilon_s^\mathrm{max}\,,
\end{equation}
where $\epsilon$ is the particle Lorentz factor, $K_s$ a normalization factor, and the extremes 
of the distribution and the power-law index will be different, depending on whether radio or optical/X-ray 
emitting particles are considered, and determined from comparison with the data. 

The local distribution function of wind particles at any place in the PWN is determined by conservation
of particle number along the streamlines, taking into account adiabatic and synchrotron losses
\begin{equation}
\label{eq:f_wind}
f_{s}(\epsilon)=\left( \frac{n}{n_0}\right)^{4/3} \left( \frac{\epsilon_0}{\epsilon}\right)^2 f_{0s}(\epsilon_0), \quad 
\epsilon_0=\left(\frac{n_0}{n}\right)^{1/3} \frac{\epsilon}{1-\epsilon/\epsilon_{\infty}} .
\end{equation}
The factors with $n/n_0$ take into account the changes in volume of the fluid element, with $n$ the total particle 
number density, continuously injected in the wind as in Eq.~\ref{eq:n},
and $n_0$ its value immediately downstream of the shock. Here $\epsilon_\infty$ 
is the maximum particle energy at every point in space-time corresponding to an injection energy 
$\epsilon_0 \rightarrow \infty$, and is evolved by the code as in \cite{Del-Zanna:2006}.

The densities $n$ and $n_0$ are evolved in space and 
time as in \cite{Camus:2009}, according to the equations
\be
\label{eq:n_n0}
\frac{\partial}{\partial t} (n \gamma) + \vec{\nabla}\cdot  (n\gamma \vec{\varv}\, )=0,
\quad
\frac{d}{dt} n_0 \equiv  \left(\frac{\partial}{\partial t}+ \vec{\varv}\cdot \vec{\nabla} \right) n_0 = 0,
\ee
where $\vec{\varv}$ and $\gamma$ are the flow velocity and Lorentz factor respectively. 
The value of $n_0$ is then used to normalize the distribution function of particles that are of wind origin 
(Case {\it A}): for these we take $K_s=\xi_s n_0$ and determine $\xi_s$ from fitting the data.
When we consider Case {\it B}, the spatially homogeneous (and constant in time) 
distribution function of radio particles is simply given as $f_R(\epsilon) = {f_0}_R(\epsilon_0)$, 
with $\epsilon_0\equiv\epsilon$ and $K_R=n_R=\mathrm {const}$ in Eq.~\ref{eq:f_0}. 

Given the local spectrum of the emitting particles, the synchrotron emissivity at all frequencies, 
surface brightness maps, and the integrated spectrum are computed through standard formulas, 
described in detail in \cite{Del-Zanna:2006}.

\section{Simulation results}
\label{sec:res}

\begin{table}
\centering
\begin{tabular}{cccc}
 \hline
 \multicolumn{2}{c}{Radio}   &     \multicolumn{2}{c}{optical/X-ray}\\
 Parameter & Value & Parameter & Value  \\
 \hline
  $p_R$      &  1.6   & $p_X$      & 2.9\\
 $\epsilon_R^\mathrm{min}$   &   $1.0\times10^3$  &  $\epsilon_X^\mathrm{min}$   & $  7.0\times 10^5$\\
 $\epsilon_R^\mathrm{max}$   &   $2.0\times10^5$  &  $\epsilon_X^\mathrm{max}$   &   $4.0\times10^9$\\
 $\xi_R$ & $1.7\times10^{-2}$ & $\xi_X$ & $2.2 \times 10^{-4}$ \\ 
 \hline
\end{tabular}
 \caption{Spectral fit parameters for the two families $f_R(\epsilon_0)$ and $f_X(\epsilon_0)$.}
\label{tab1}
\end{table}

This section is devoted to comparison of the radio emission morphology resulting from the 
different assumptions on low-energy particle acceleration illustrated above.
We have performed several numerical simulations aimed at determining what values 
of the dynamical parameters lead to emission maps that most closely reproduce the Crab Nebula 
morphology at high energy. With respect to previous work on the subject \citep[see][]{Del-Zanna:2006,Volpi:2008}, 
we have explored a much wider portion of parameter space, including values of $\sigma_0$ 
larger than unity, and also varying, for the first time to our knowledge, the wind anisotropy 
parameter $\alpha$. Detailed results of these simulations will be presented elsewhere, 
while for the current purpose what is important is the identification of the flow parameters 
that provide the best description of the Crab Nebula X-ray morphology. 
Even if we employ a different simulation set up, the best set of values for the free 
parameters is still that found by \cite{Del-Zanna:2006}: $\alpha=0.1$, $\sigma_0=0.025$, $b=10$.
The parameters describing the particle spectrum at the basis of the emission 
are reported in Table~\ref{tab1}, and will be discussed later on.

\begin{figure}
\includegraphics[scale=.5]{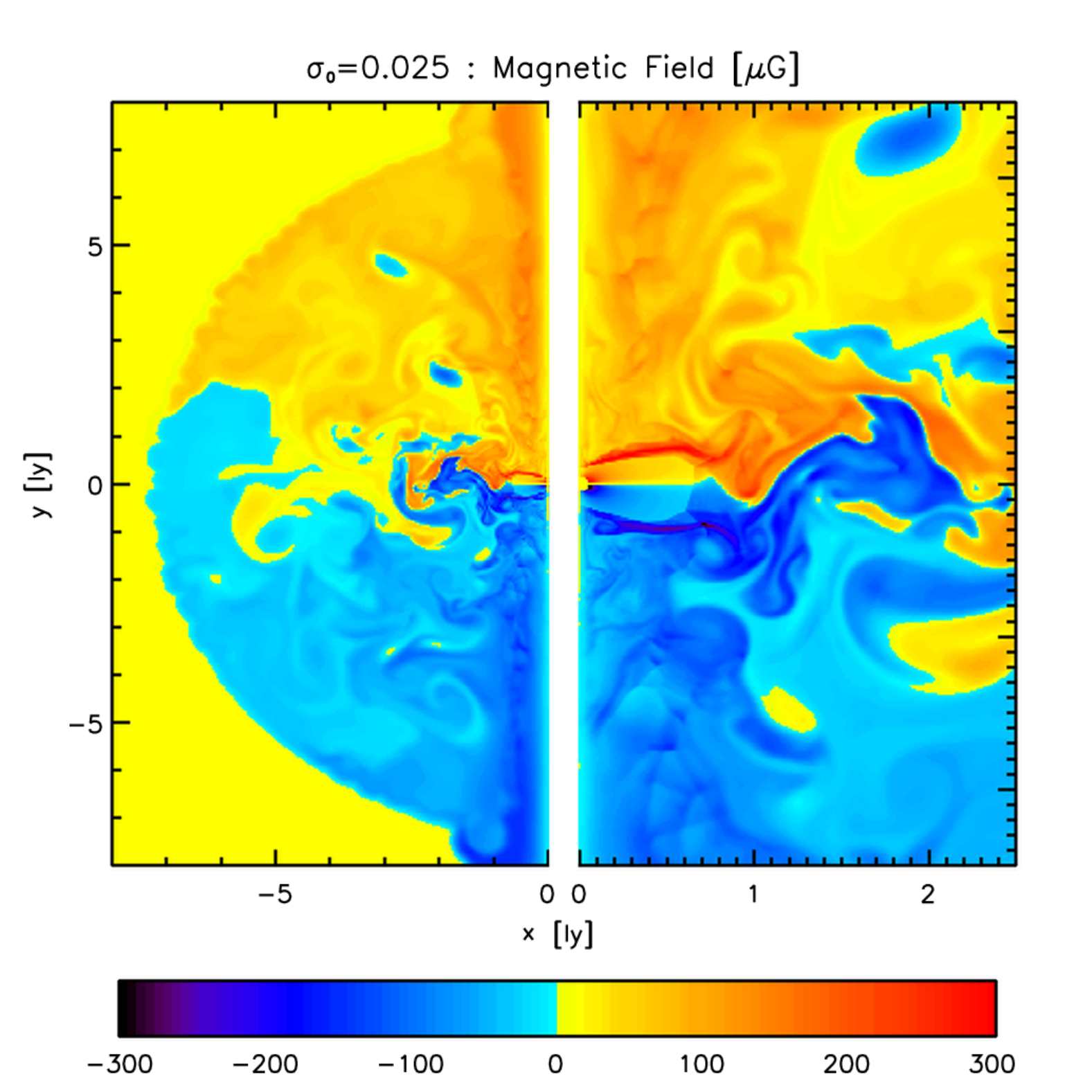}
\caption{The magnetic field distribution in the simulated nebula.} 
\label{fig:bphi}
\end{figure}

\begin{figure*}
  \includegraphics[scale=.35]{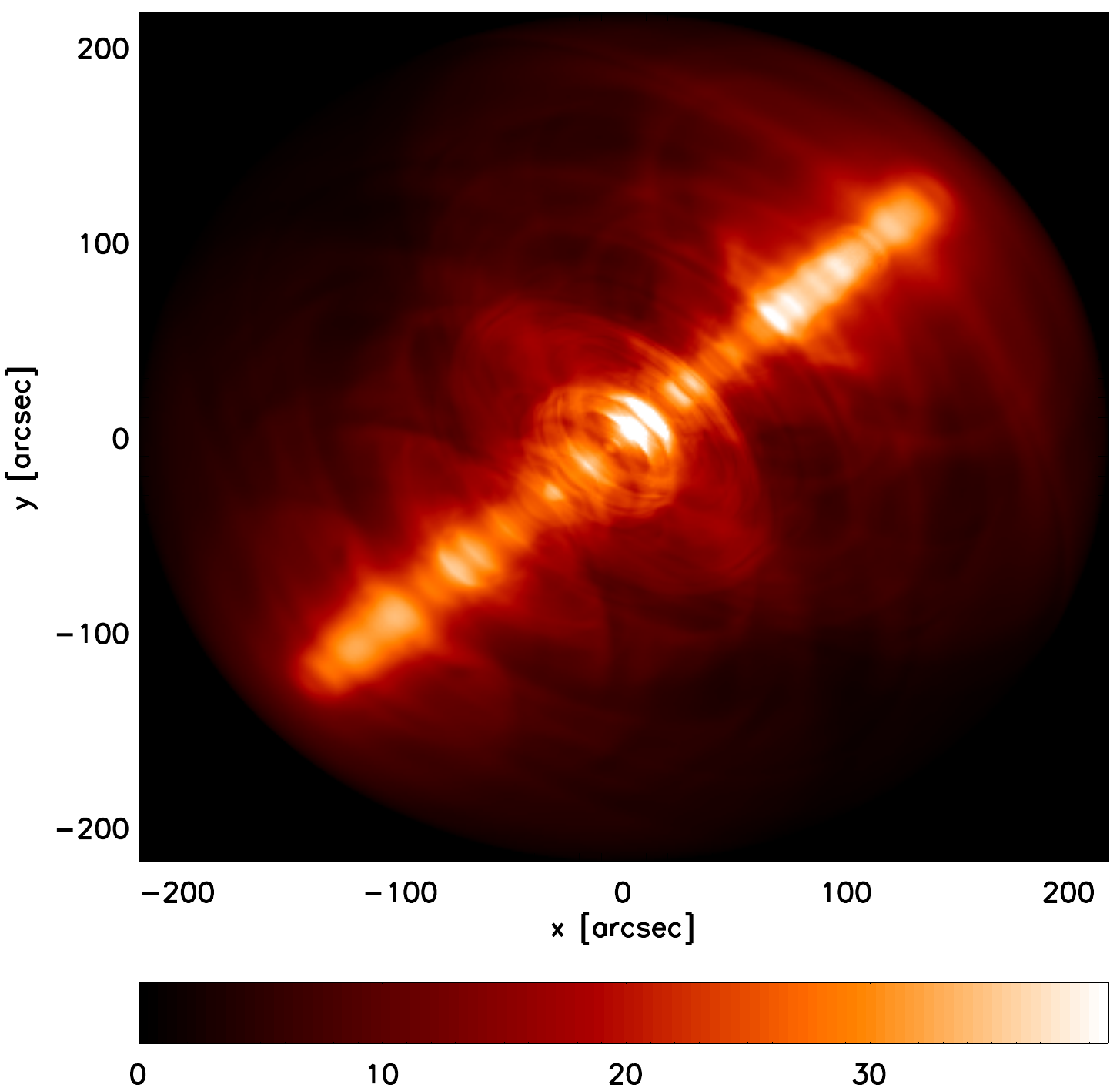}
  \includegraphics[scale=.35]{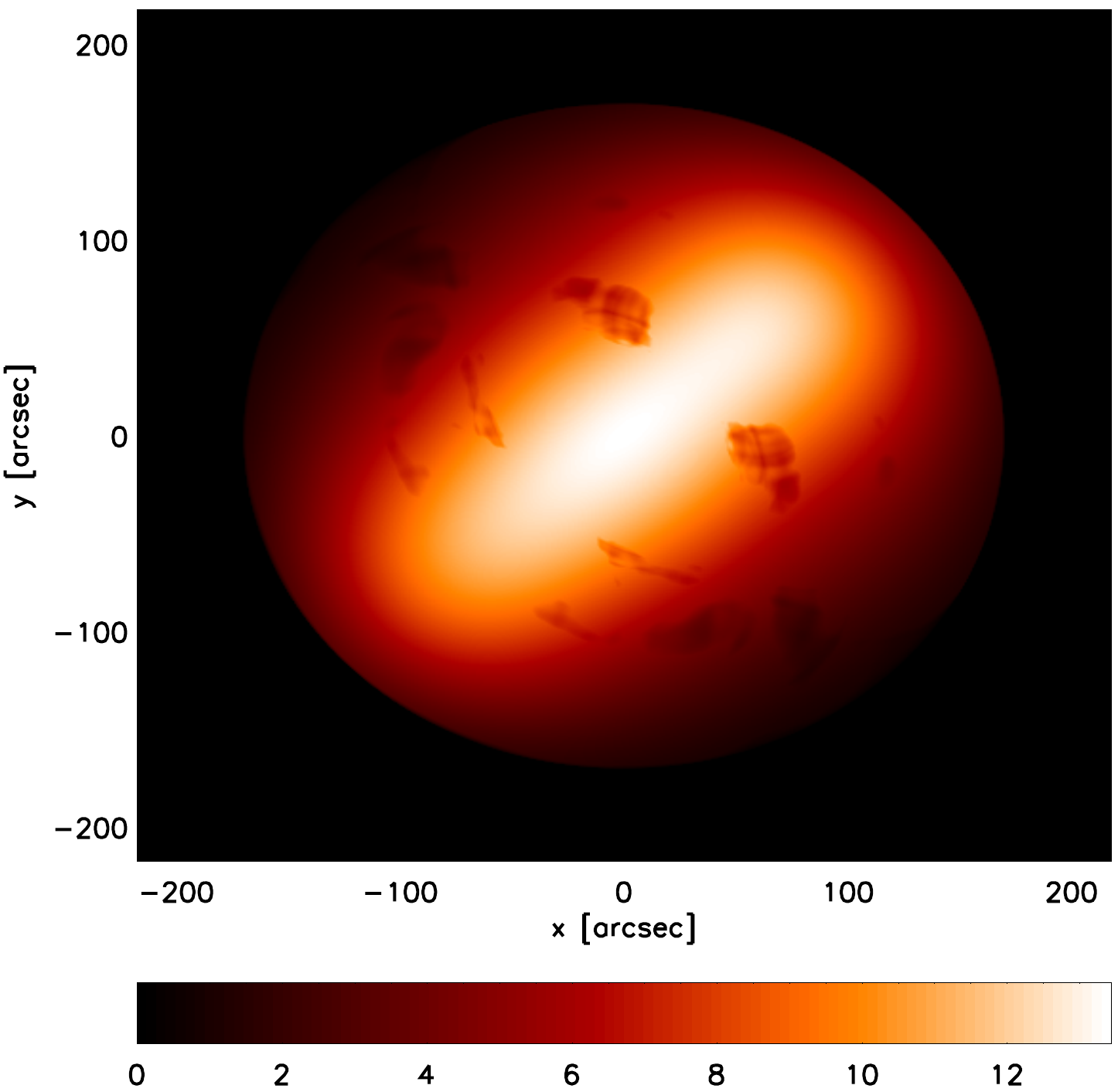}
 \includegraphics[scale=.35]{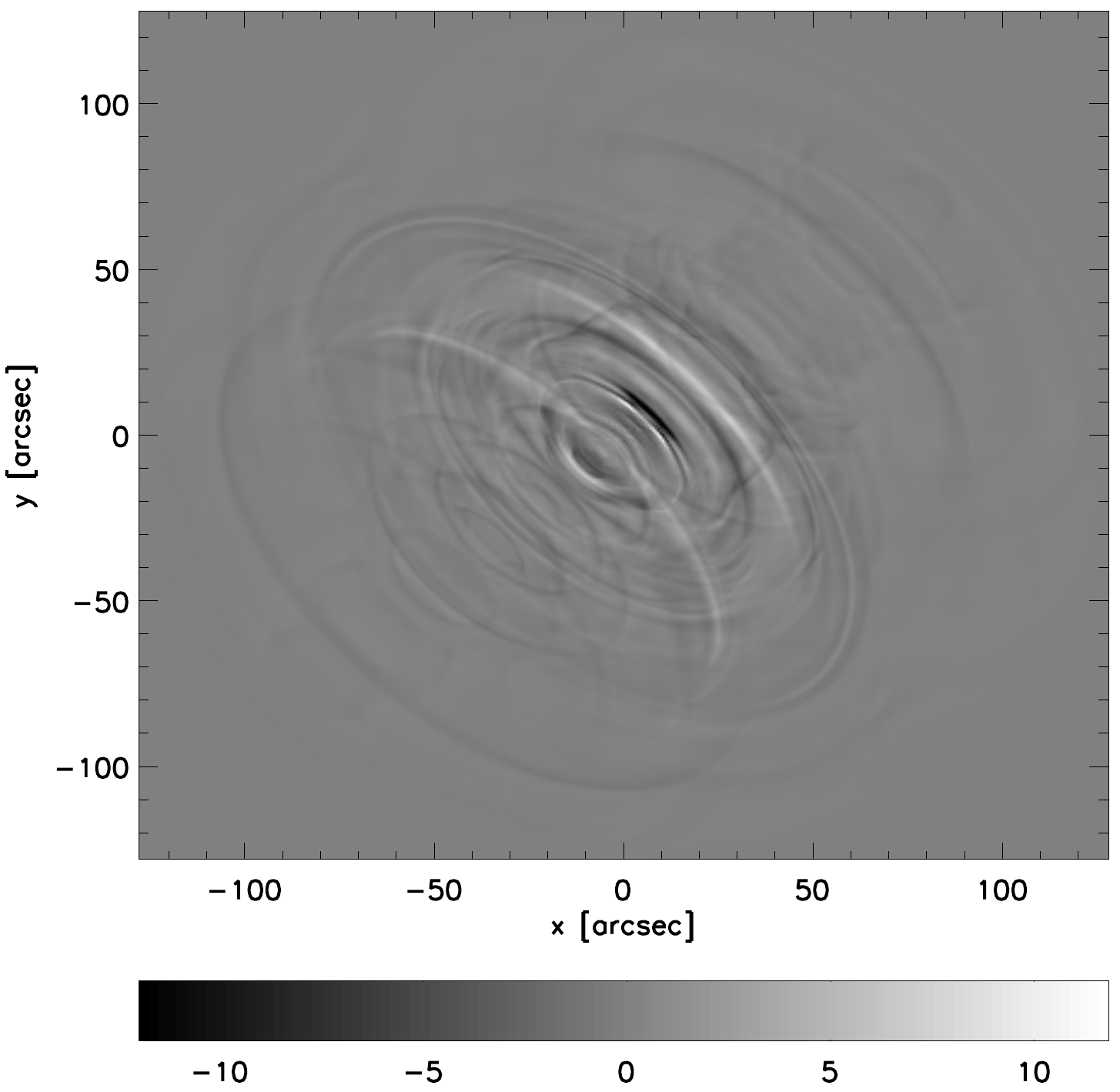}\\
 \includegraphics[scale=.35]{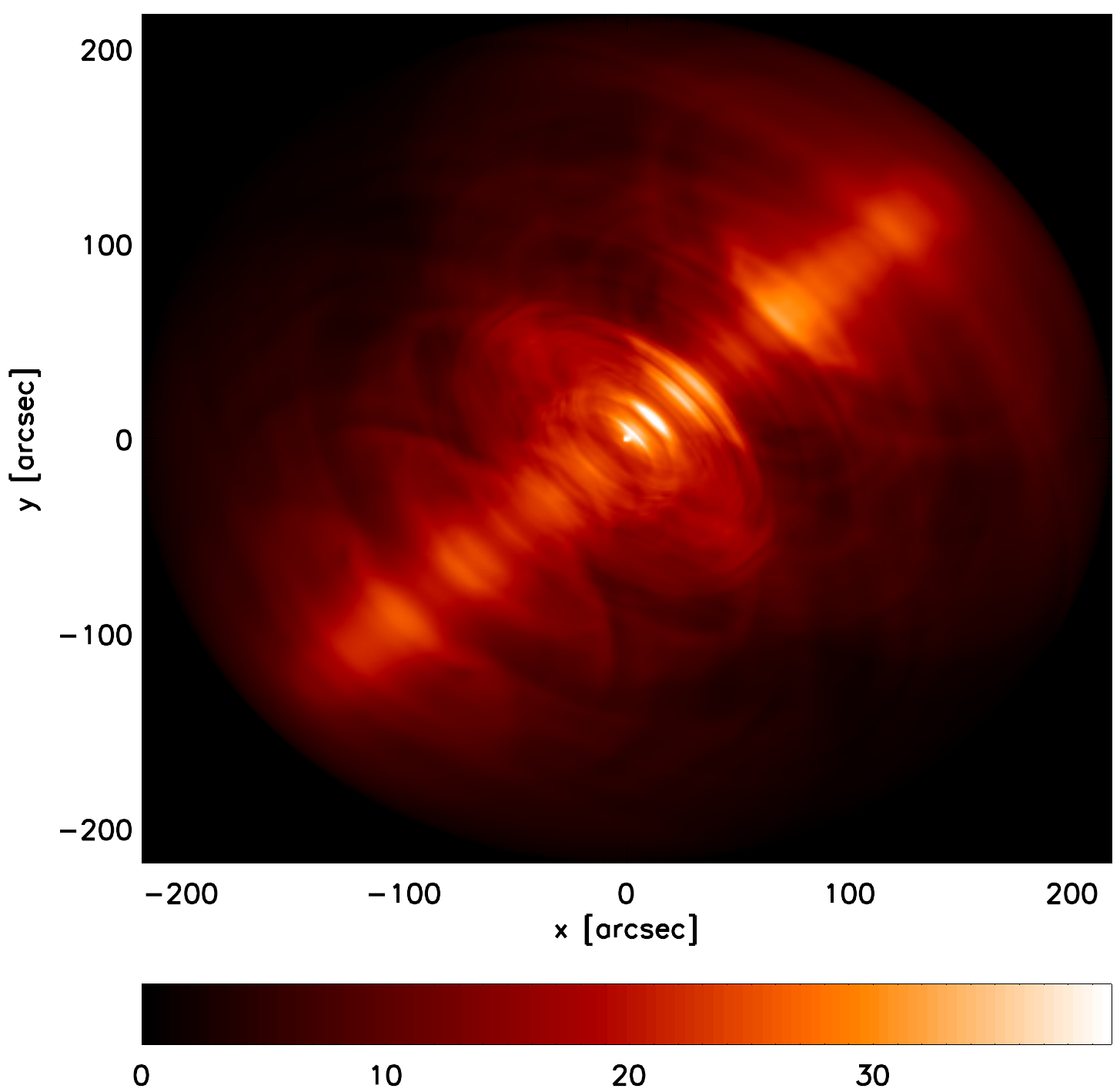}
   \includegraphics[scale=.35]{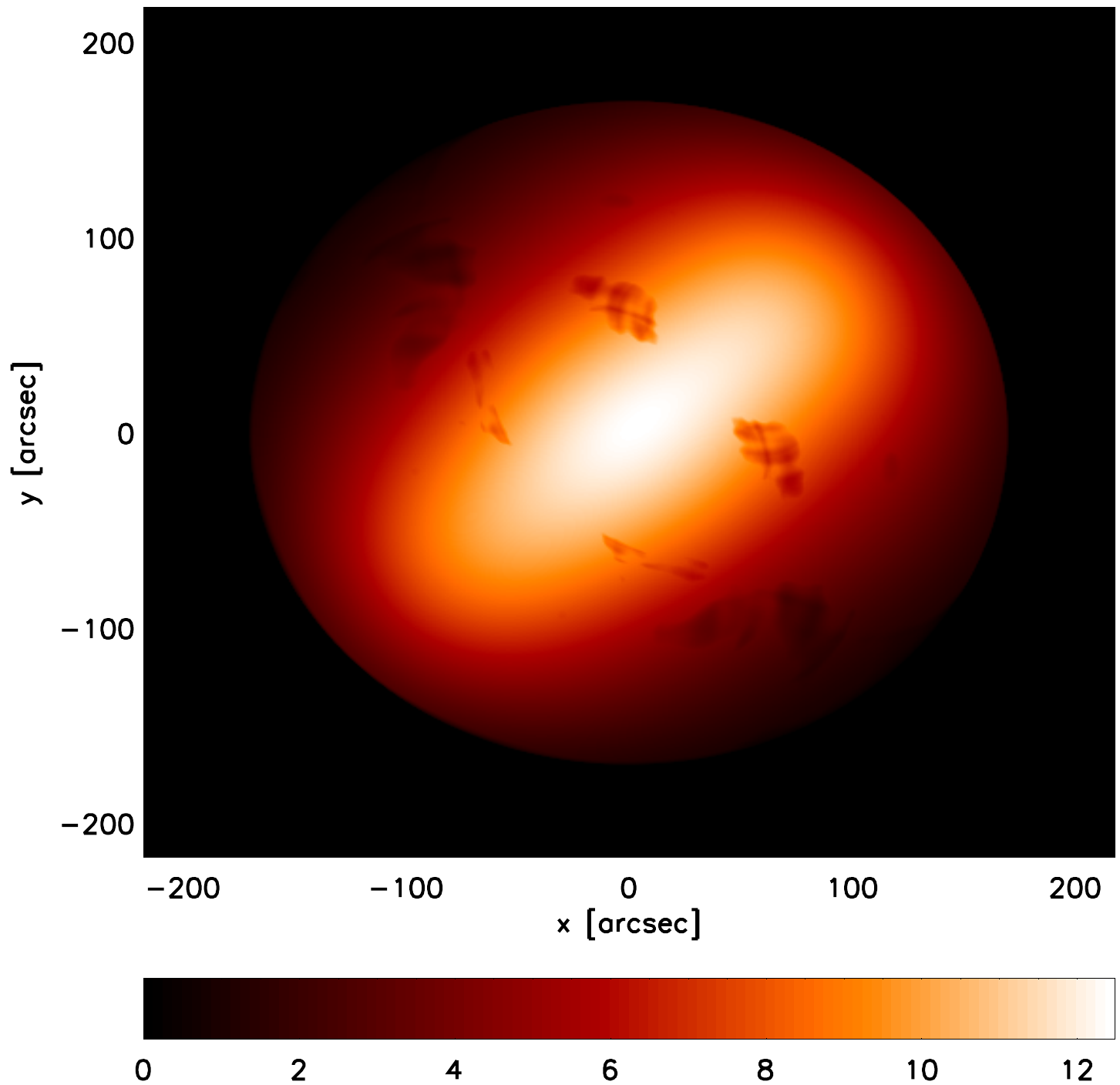}
 \includegraphics[scale=.35]{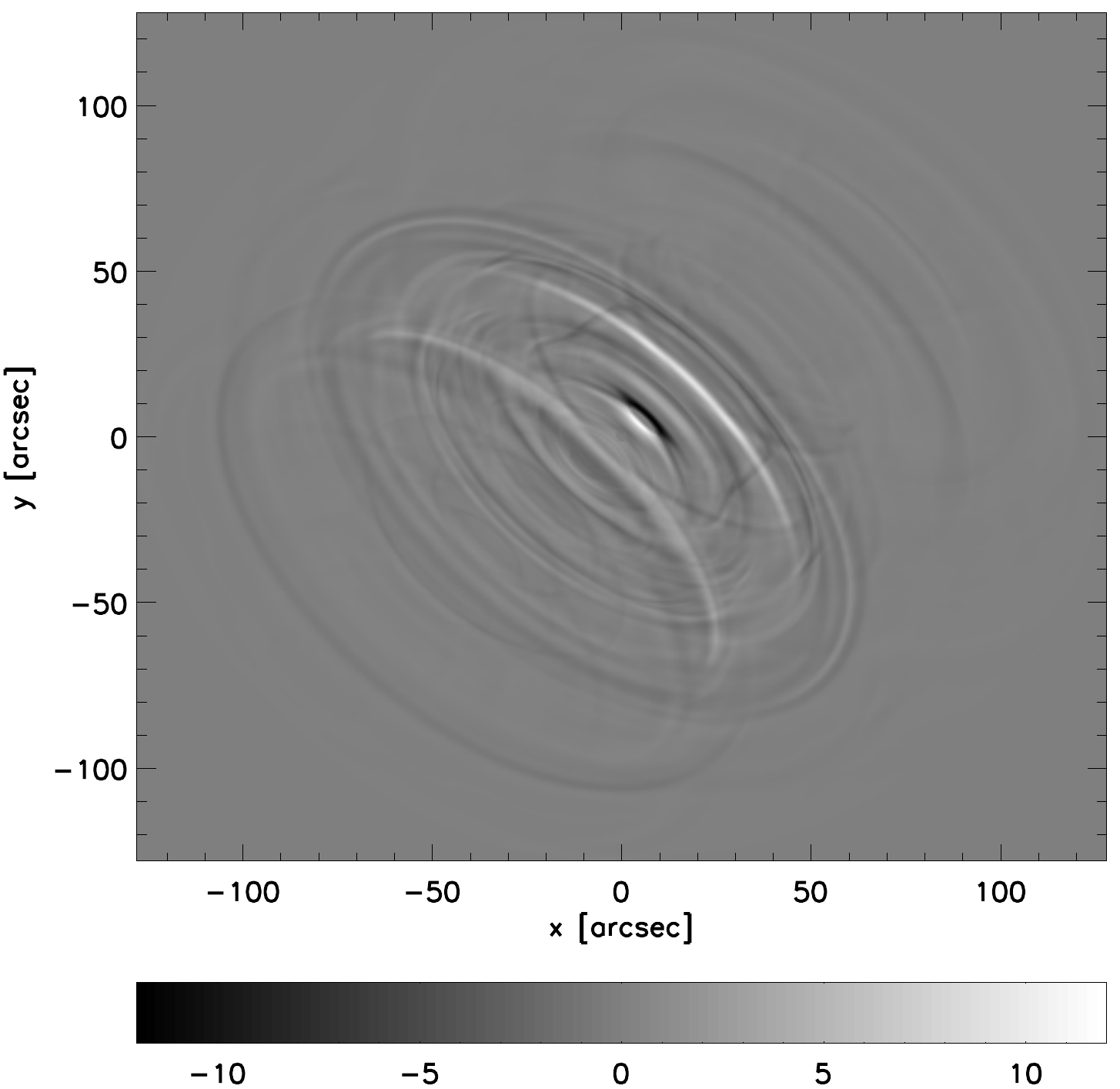}
\caption{Simulated maps of radio emission and \emph{wisps} at 1.4 GHz, expressed in mJy/arcsec$^2$. 
\textit{Left panels}: surface brightness maps; \textit{middle panels}: surface brightness maps with 
subtraction of small scales; \textit{right panels}: \emph{wisps} obtained by subtracting two surface 
brightness maps with a time separation of two months. The upper row refers to Case {\it A} 
and the lower one to Case {\it B}, for which $n_R=1.4 \times 10^{-6}{\rm cm}^{-3}$.}
 \label{fig:mapR}
\end{figure*}

In Fig.~\ref{fig:bphi} we show the (toroidal) magnetic field at $t=950$~yr of evolution. 
The size of the termination shock ($\simeq 0.7$~ly at the equator) and that of the contact discontinuity 
with the expanding ejecta ($\simeq 7$~ly) are approximately as observed. 
The field changes polarity around the equator and eddies cause the current sheet to twist 
and tangle in the downstream, creating a strong mixing of the field 
(\emph{islands} of opposite polarity can also be seen) and consequent dissipation \citep{Bucciantini:2013}. 
The field is lower than expected basically everywhere:
well below $10^{-4}$~G in the external regions and $\simeq 2\times10^{-4}$~G just
downstream of the termination shock and within a cylindrical region of radius $\sim 1$~ly around the 
polar axis (where a jet is present due to magnetic hoop stresses).

\subsection{Shock versus distributed particle acceleration}

In this subsection we compare the radio emission morphology resulting 
from assumptions {\it A} and {\it B}, namely ongoing acceleration of the low energy particles 
at the termination shock, followed by advection with the flow (in perfect analogy with optical/X-ray 
emitting particles, Case {\it A}), or uniform distribution in the whole nebular volume (Case {\it B}).

\begin{figure*}
 \includegraphics[scale=.55]{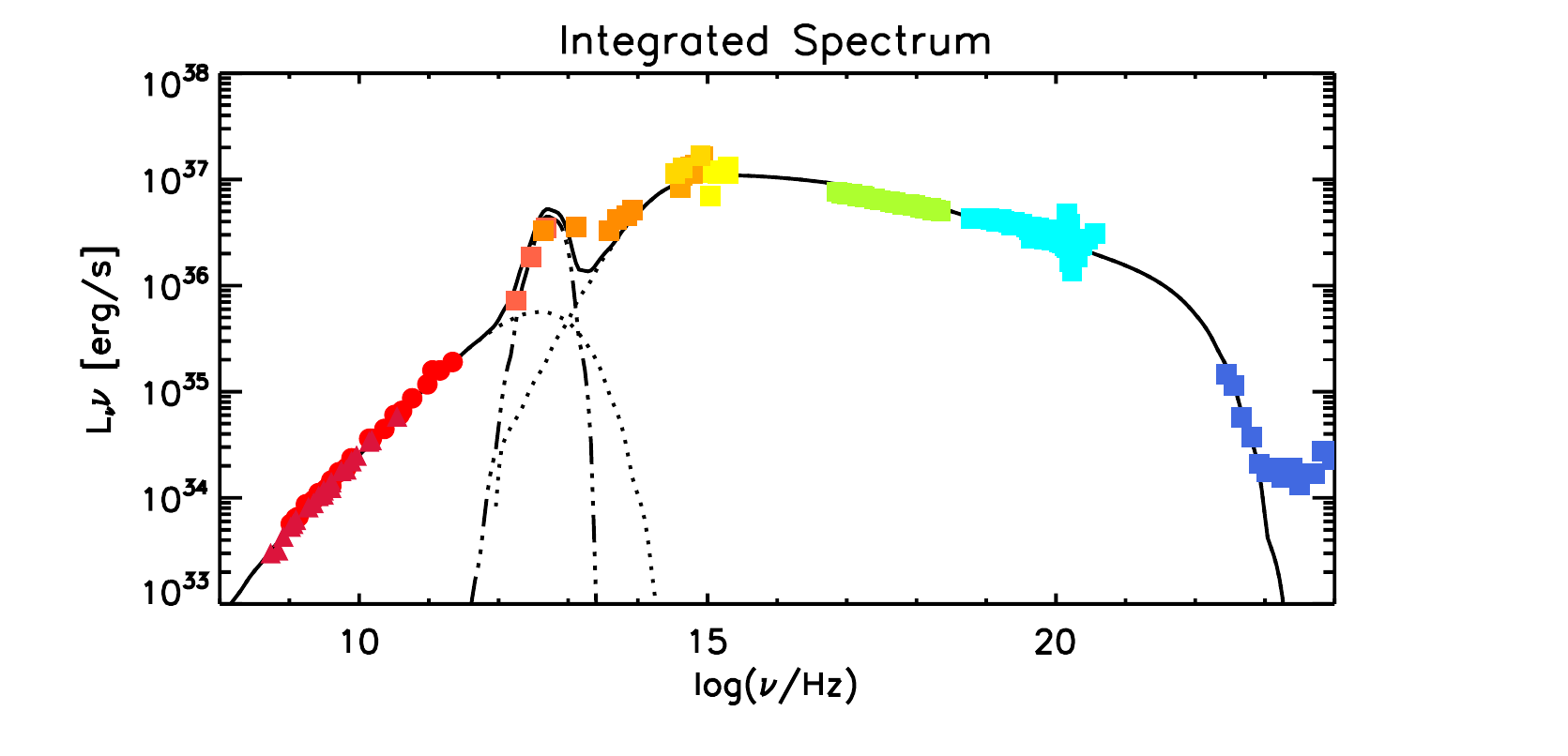}
 \hspace{-10mm}
 \includegraphics[scale=.55]{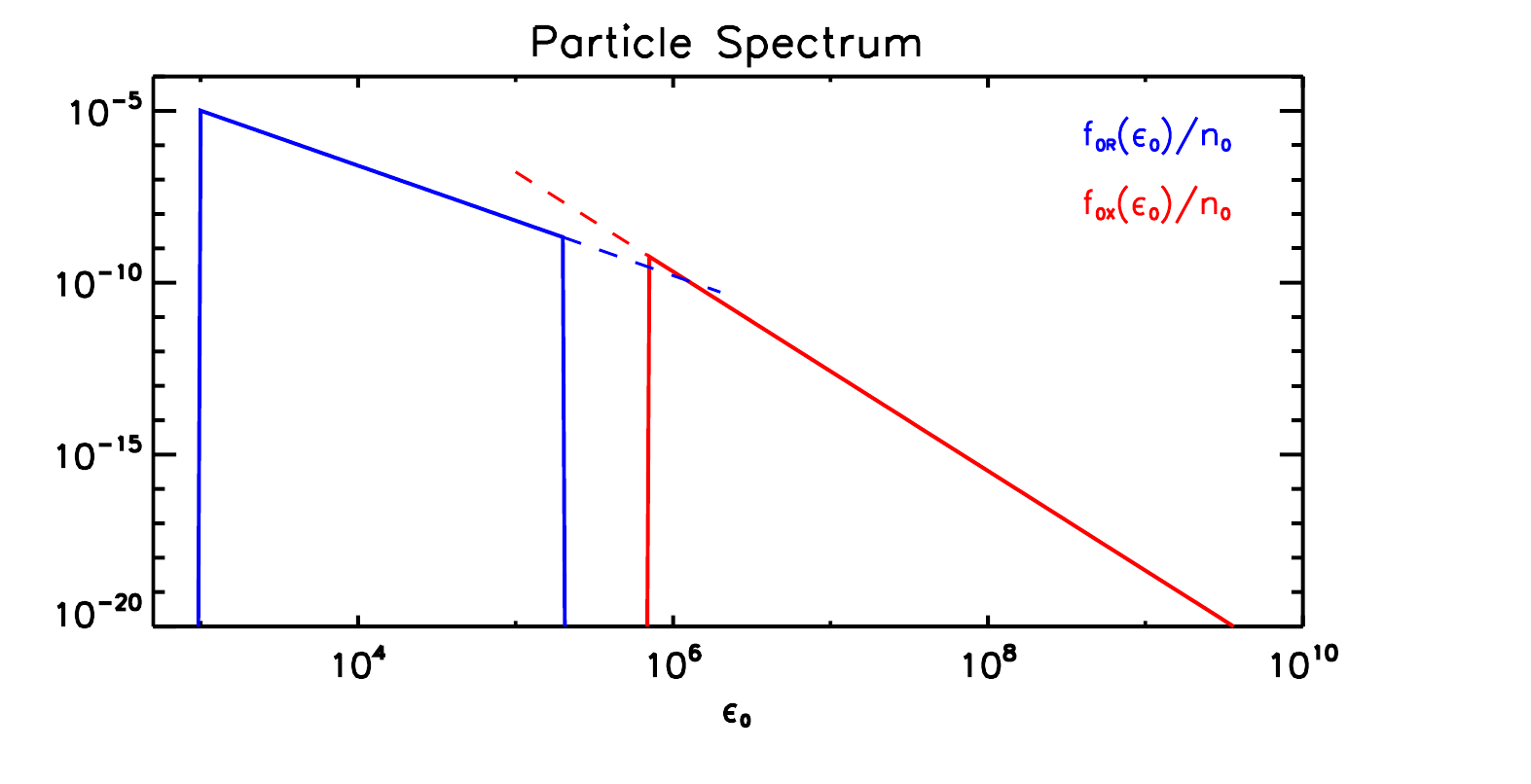}
 \caption{The Crab Nebula integrated synchrotron spectrum. Data are taken from \citet{Meyer:2010}. 
 Dotted lines indicate the contributions from the two families of emitting particles, whereas the
 solid one represents the total spectrum.
 The injection spectra are shown in the right panel for the case of radio 
 particles of wind origin (Case {\it A}). The blue (radio particles) and red curve (optical/X-ray) 
 correspond to the parameters in Table~\ref{tab1}.
 }
  \label{fig:spec}
\end{figure*}

The radio emission maps, which are here computed for the first time on top of MHD simulations, 
directly reflect the magnetic field structure, and, at some level, the structure of the velocity field 
(due to Doppler boosting). In the top row of Fig.~\ref{fig:mapR} we show the
results of Case {\it A}, whereas the lower row pertains to Case {\it B}. In the left panels we report
the surface brightness maps at 1.4~GHz. The maps computed under the two different assumptions 
on the origin and distribution of radio emitting particles are basically identical. 
Ring-like structures, very similar to those observed in the optical and X-rays are clearly 
seen at these long wavelengths too. Their appearance is in fact only due to local enhancements 
in the magnetic field and Doppler boost associated with the local velocity field. In the right panels 
we show images obtained by subtracting two maps computed for epochs that are two months apart: 
\emph{wisp}-like motion is clearly seen also at radio wavelengths, in perfect agreement with what observed 
by \cite{Bietenholz:1997}. Again, no substantial difference between the two cases {\it A} and {\it B} can
be noticed, the only discrepancy being due to the central slightly brighter features.
Our conclusion is that the appearance of radio \emph{wisps} does not bear any implication on 
the injection site of the emitting particles, but only on the underlying flow structure.

On the other hand, the emission structure we find appears to be too strongly concentrated 
in the axial direction in comparison with observations: this is again independent on the assumed 
spatial distribution of particles in the nebula. One could argue that such an effect is actually 
present in the real data but masked by the presence of thermal filaments, which give a 
non-negligible contribution in this band. We compared our results with the map at 1.4 GHz 
published by \cite{Bandiera:2002}, where subtraction of the filaments was performed.
In the central panels of Fig.~\ref{fig:mapR} we show our emission maps smoothed with the
same technique and using the same units of mJy/arcsec$^2$ as in the cited work for direct comparison. This smoothed maps appear to be very similar to the one based on observations from both a qualitative and quantitative point of view. The computed emission
shows a strong cylindrical symmetry, somewhat at odds with the ellipsoidal structure in the data. 
Once more, results for Case {\it A} and {\it B} are nearly undistinguishable.

The reason for the latter discrepancy is likely related to the fact that axisymmetric simulations 
do not provide a good description of the magnetic field on large scale. 
Kink-type instabilities that are likely to be at work in these nebulae \citep{Begelman:1998}, 
are artificially suppressed in 2D. This might reflect in a simulated magnetic field that 
has a much higher degree of order than in reality: the discrepancy will become more 
apparent in the emission the further one moves from the central pulsar, so that the instabilities 
have time to grow. This fact makes the radio emission the most sensitive probe of such an effect, 
and at the same time it explains why with axisymmetric simulations we are able anyway to 
reproduce the high energy morphology: the short lifetimes of optical/X-ray emitting particles 
keeps them confined within a region where a perfectly toroidal magnetic field is still not such 
a bad approximation. An additional hint to such an effect comes from the observation 
that in order to reproduce the observed brightness contrast between 
the inner ring and torus at X-ray frequencies the emission must be computed 
assuming that the underlying magnetic field progressively becomes isotropic 
with increasing distance from the termination shock: the ring is otherwise too bright.

The artificially high degree of order of the magnetic field is obviously accompanied 
by important dynamical consequences. The associated hoop stresses are also larger 
than in reality and might be forcing us to adopt an artificially small value of the 
magnetization in our simulations. Strong suggestions that this is the case also come 
from Fig.~\ref{fig:spec}, where we plot the integrated emission spectrum of the Crab Nebula 
as computed based on our simulation, and we compare it with multi-wavelength 
observations. The particle distribution functions adopted are as described in 
Sec.~\ref{sec:setup} with the parameters given in Table~\ref{tab1}. 
In spite of the good fit of the synchrotron part of the spectrum, the 
exceedingly low magnetic field strength in the nebula is revealed by a careful
inspection of the fit parameters: the spectrum of optical/X-ray 
emitting particles that is required to reproduce the observations is very steep (see Table~\ref{tab1}), 
with a slope of $p_X=2.9$, to be compared with the value $p_X=2.2$ that is deduced from 
observations of the inner nebula \citep{Veron-cetty:1993,Mori:2004}. 

This is due to the fact that synchrotron losses are too weak in our simulation 
and do not provide sufficient softening of the injected particle distribution, 
so that we are forced to assume a steep injection spectrum in order to reproduce 
the high energy observations.
However, this becomes even more apparent from the gamma-ray flux
due to the IC contribution, which, when calculated self-consistently by
integrating over the whole nebular volume, is higher by a factor $\sim 3$
compared to observations. 
This is precisely the discrepancy we have in the average
magnetic field value: in order to reproduce the synchrotron spectrum
we are forced to inject more particles in the PWN, leading to a higher IC flux. 
This was already found and described in details by \cite{Volpi:2008}:
very similar findings apply to the simulations with the present settings.

Finally, let us comment on the properties of the particle distribution function underlying 
the emission, focusing on Case {\it A}. This case corresponds to the spectrum described in 
Table~\ref{tab1} and Fig.~\ref{fig:spec} (right panel). 
First thing to notice is that the spectrum of emitting particles we adopt has a gap
between the radio and the X-rays. The size of the gap is not well established from our fit: 
anything between no discontinuity ($\epsilon^\mathrm{max}_r=\epsilon^\mathrm{min}_X$) 
and a discontinuity extending to an energy a factor of 3-4 lower than 
$\epsilon^\mathrm{min}_X$ works well (the ratio for our best fit is 3.5). This result is
obtained with sharp cut-offs in the particle spectra both at high and low energy; 
while with exponential cut-offs larger gaps would be allowed.
In addition, our best fit corresponds to a discontinuous distribution at injection, in the sense
that the two curves do not match in the gap region and 
that extrapolation of the low energy spectrum to $\epsilon^\mathrm{min}_X$ would give half 
of the particles that are actually required at this energy. 
However, this discrepancy should not be trusted because, as already discussed,
inference of the particle spectrum depends on the magnetic field strength and profile and we 
know that in reality this will be different than in our current axisymmetric simulations. Indeed,
the spectra described in Table~\ref{tab1} correspond to conversion into accelerated
particles of a fraction larger than one of the pulsar spin-down luminosity.

\subsection{On the variability of radio ``wisps''}
\label{subsec:radio-em}

\begin{figure*}
 \includegraphics[scale=.4]{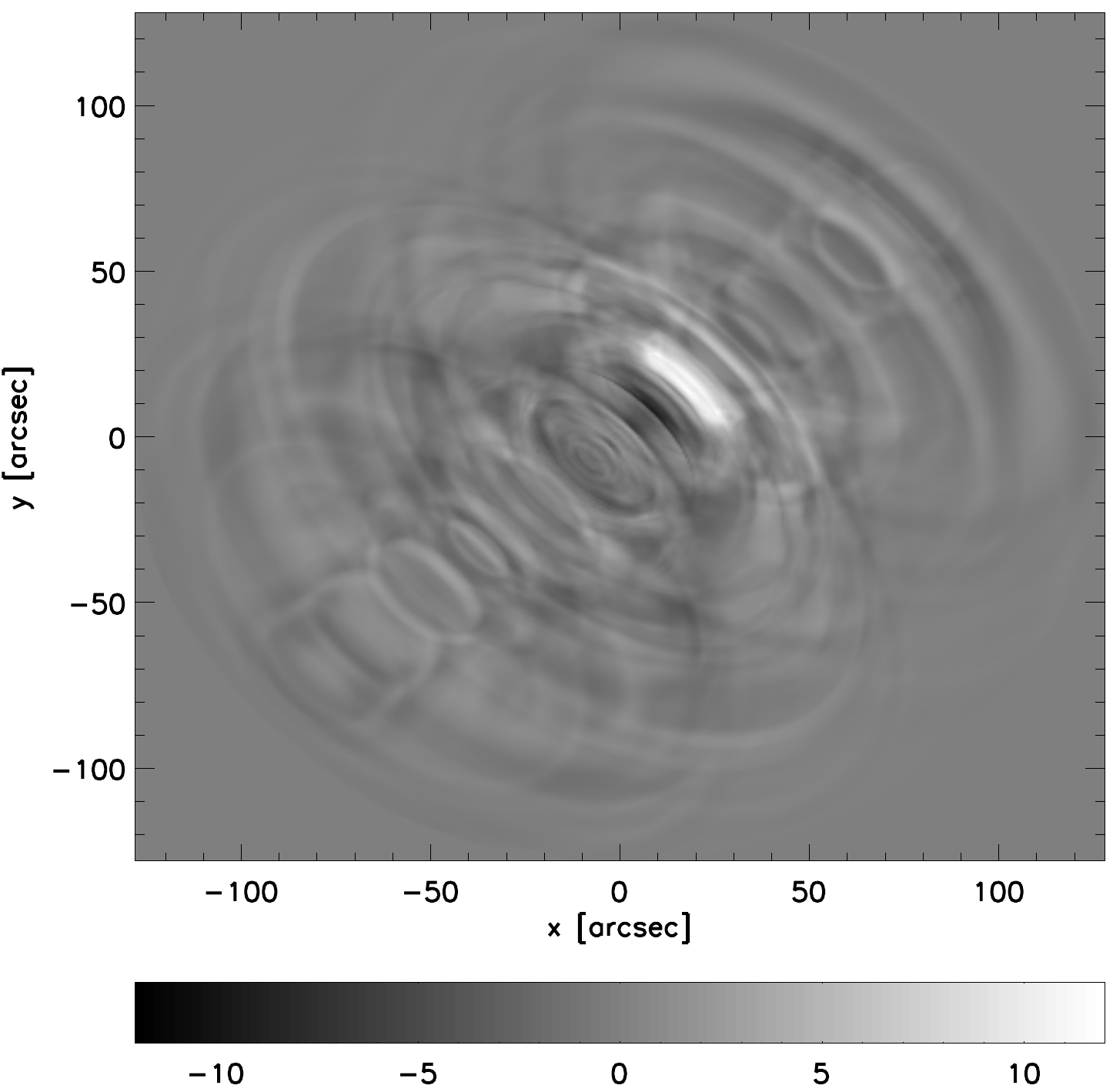}
 \hspace{-4mm}
 \includegraphics[scale=.4]{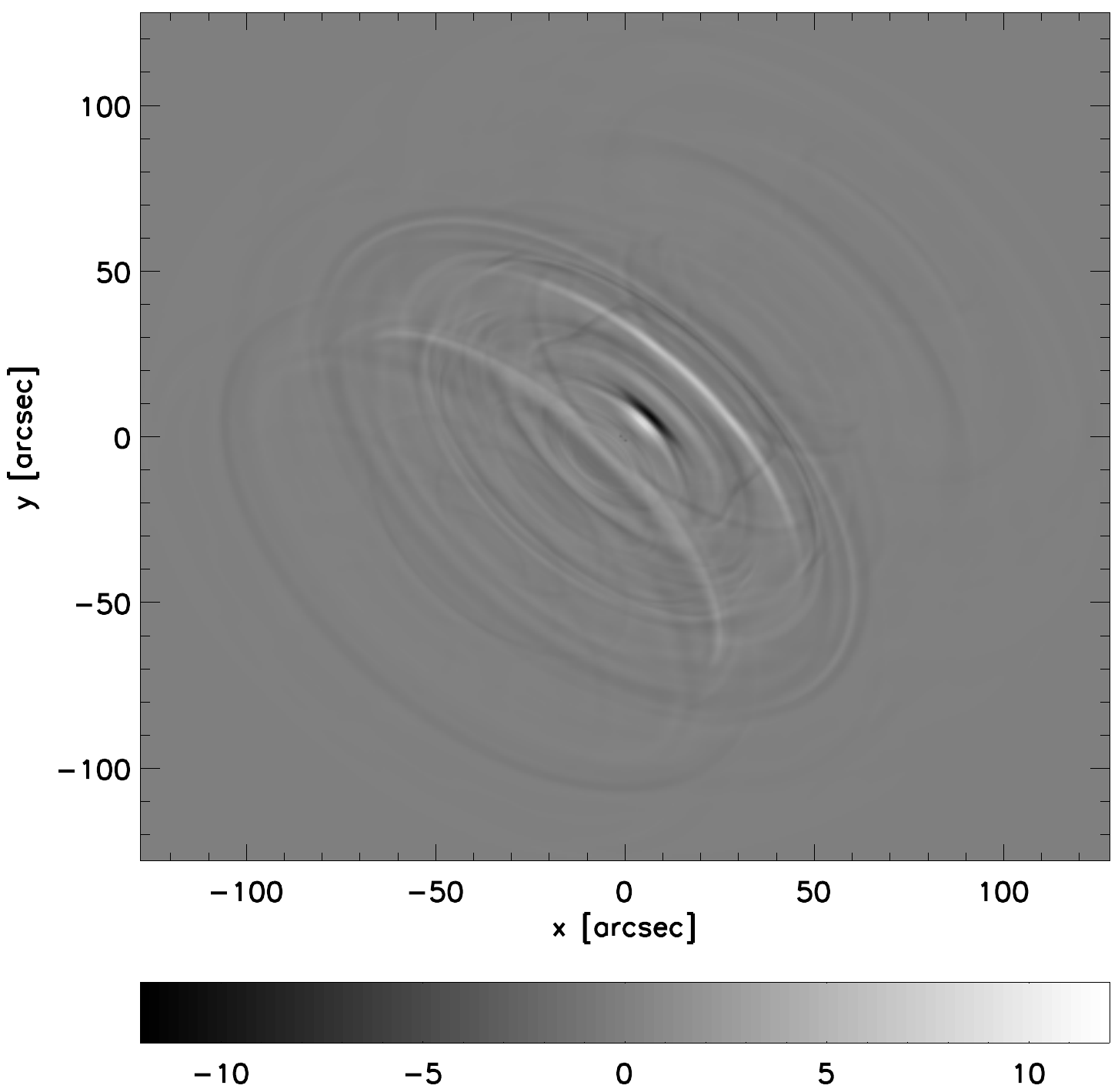}
 \caption{Changes occurring in the nebula on a 3 yr time-scale ({\it left panel}) and on 
 $\sim 2$ months ({\it right panel}), at 5 Ghz. Moving ripples and arc-like structures are well 
 visible in both images. On the $x$ an $y$ axes the distance from the central pulsar 
 in arcsec is reported. The two maps are expressed in mJy/arcsec$^2$.}
  \label{fig:radio-var}
\end{figure*}

We have seen in the previous subsection that radio \emph{wisps} are easily reproduced
by our relativistic MHD model. Here we investigate their variability properties following the
analysis by \cite{Bietenholz:2004}. These authors studied long-term variability of the radio emission 
component through observations during a $\sim 3$~yr period, together with much shorter-term
variations (time-scale of a few months) in the innermost region of the nebula.
For direct comparison with this analysis of time-variability, we use maps that are computed at 5 GHz, 
instead of the previously used frequency of 1.4 GHz. As above, the difference images are obtained
by subtraction.
We only discuss maps corresponding to Case {\it B}, since those for Case {\it A} are very similar.

In Fig.~\ref{fig:radio-var} the radio emission difference maps at 5 GHz are shown: the long-term 
is shown in the left panel, short-term in the right one. As one can expect, the right panel corresponds to the radio 
{\it wisp}-like structures in the inner region of the nebula, just as in the maps at 1.4 GHz already shown 
(images on the right in Fig.~\ref{fig:mapR}).
As far as the long-term map is concerned, the first thing to notice is that a significant component 
of the variability has a time-scale longer than that of the {\it wisps}. The region that shows long term  
variability extends for $\sim 3$ ly around the pulsar, namely it is twice as large in linear size as the 
region showing short-term variability.  
Additional remarkable features in the image on the left of Fig.~\ref{fig:radio-var} are: elliptical arcs, 
found in the central zone and characterized by an outward motion, similar to the optical/X-ray {\it wisps}; 
arc-shaped features that are visible in the body of the nebula, are not centered on the pulsar position and 
move outward more slowly than the former \citep{Weisskopf:2012,Schweizer:2013}. 
The inferred projected speed for the inner \emph{wisps} is $\sim 0.4c$ while for the outer moving features 
is $\sim 0.002c$, as expected from the data analysis.

All these features were found in the data analysis by \cite{Bietenholz:2004} and are well reproduced by 
our axisymmetric model both in shape and associated velocity. We stress again the conclusion that the 
appearance of \emph{wisps} simply reflects the underlying structure of magnetic and velocity field 
(due to Doppler boosting). The result is not much affected by the different spatial distribution of the 
particles in the 2 cases {\it A} and {\it B}: indeed, even in the case of acceleration at the shock and 
following advection ({\it A}), thanks to their long lifetimes, the radio particles are everywhere in the 
Nebula and strong gradients are absent.

\subsection{Pure ``relic'' particles: no re-acceleration}
\label{subsec:tdep}

In this section we discuss what we had defined as our Case {\it C}. We consider a model in which radio emitting 
particles are pure relics: they are injected at the termination shock (as the optical/X-ray emitting one) 
but only at early times after the SN explosion. The injection of low energy particles is then halted after 
a certain time $t_\mathrm{b} \ll t_\mathrm{act}\simeq 950$ yr, while continuing for the the higher energy ones. 
After injection all particles evolve due to synchrotron and adiabatic losses, while advected with the flow. 
This time-dependent injection model is meant to represent a sort of primordial \emph{burst}, possibly 
associated to a different {\it behavior} of the pulsar in its early stages, when the spin-down luminosity 
was almost an order of magnitude greater than the current one \citep{Atoyan:1999}: what we have 
in mind is in fact a higher multiplicity at early times. Our attention is then focused on the final spatial 
distribution of the radio particles and on the resulting emission map. 

Also for the present Case {\it C} the simulation set-up is the same as before, except for the particle tracers that are now defined only for the radio population, with the new value of the maximum local energy set as ${\epsilon_\infty}=10^6$.

This value is chosen in order to guarantee a proper representation of the radio emission spectrum. 
In Sect.~\ref{sec:res} we found that the radio spectrum is currently produced by particles with energy up 
to a cut-off $\epsilon_\mathrm{b}\simeq 2 \times 10^5$. The most recent radio particles were injected in the nebula 
at the time $t_b$ corresponding to the end of the burst. Since then they have been losing energy due to adiabatic, 
and, initially, also synchrotron losses: at early times the average magnetic field in the nebula was larger 
than its current value making synchrotron losses important also for $100$ GeV particles. 
This fact actually constrains the burst duration: if we want the current spectrum of radio emitting particles 
to still extend up to the value of $\epsilon_\mathrm{b}$ mentioned above, we require that the burst must 
have lasted at least for a time:
\begin{equation}
t_\mathrm{b} \gtrsim \sqrt{\frac{\epsilon_\mathrm{b} \sigma_T c t_\mathrm{act}^3 
B_\mathrm{act}^2}{18 \pi m_\mathrm{e} c^2}}\sim 80 \left(\frac{\epsilon_\mathrm{b}}{2 \times 10^5}\right)^{1/2}
\left(\frac{B_\mathrm{act}}{50 \mu G}\right) {\rm yr}\ ,
\label{eq:tb}
\end{equation}
where we have used standard synchrotron formulae and approximated the time evolution of the 
average magnetic field strength as $B(t)\propto t^{-1}$ \citep{Pacini:1973}. Since the average magnetic field 
in our simulation at the current age of the Crab Nebula is $B_\mathrm{act}\sim 50 \mu G$, we take $t_\mathrm{b}=100$ yr.

We would like to point out however, that if the magnetic field in our simulation were closer to more 
realistic values (values in agreement with high energy spectral modeling), and hence a factor 2-5 larger, 
the lower limit on the burst duration would have been proportionally larger. Therefore, in reality, 
the highest energy radio emitting particles must have been injected in the nebula no earlier than $\sim$ 500 yr ago, which makes one wonder what might have caused a sudden change in the pulsar multiplicity or 
spin-down properties long after birth.

\begin{figure*}
 \includegraphics[scale=.35]{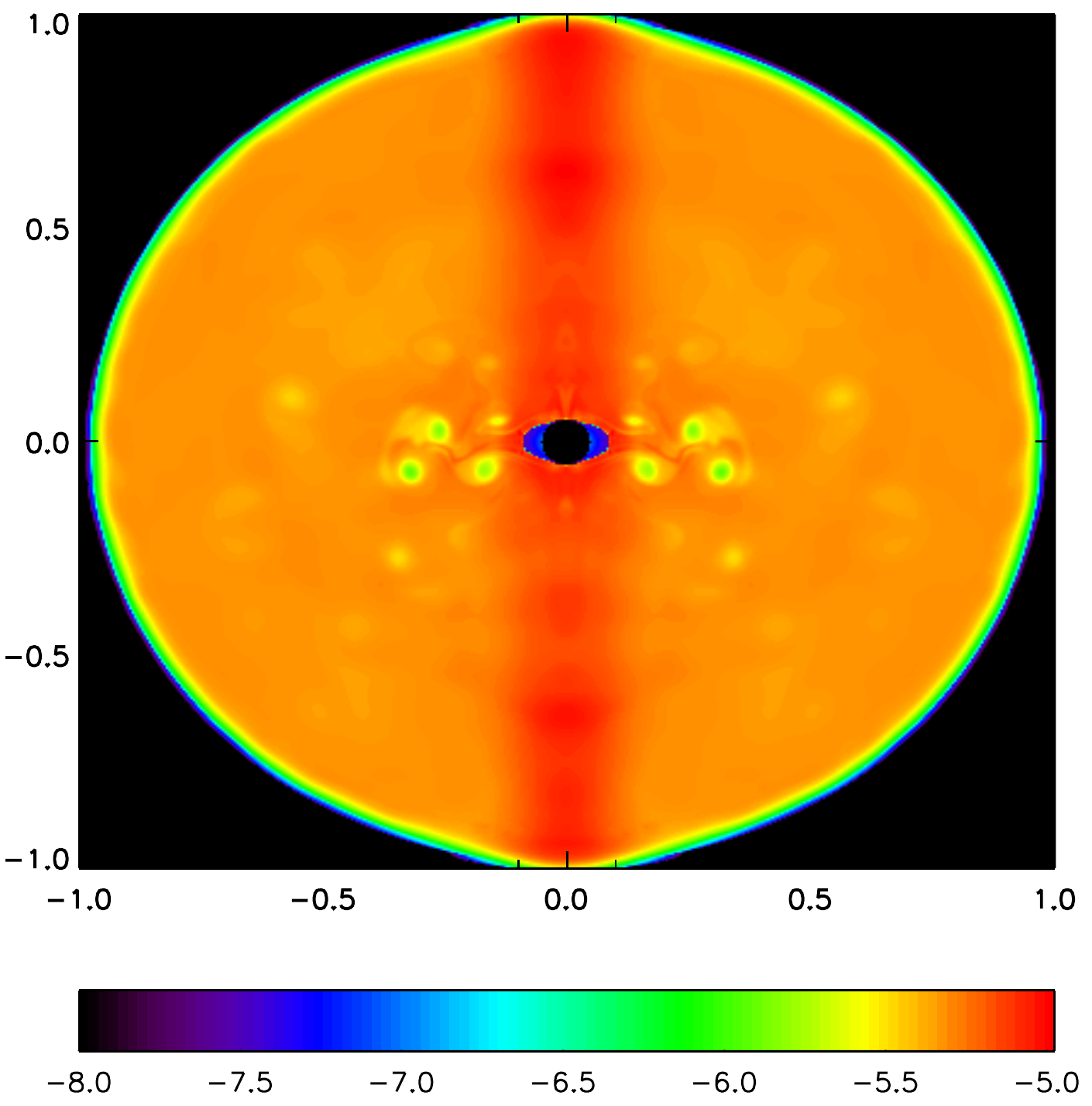}
 \includegraphics[scale=.355]{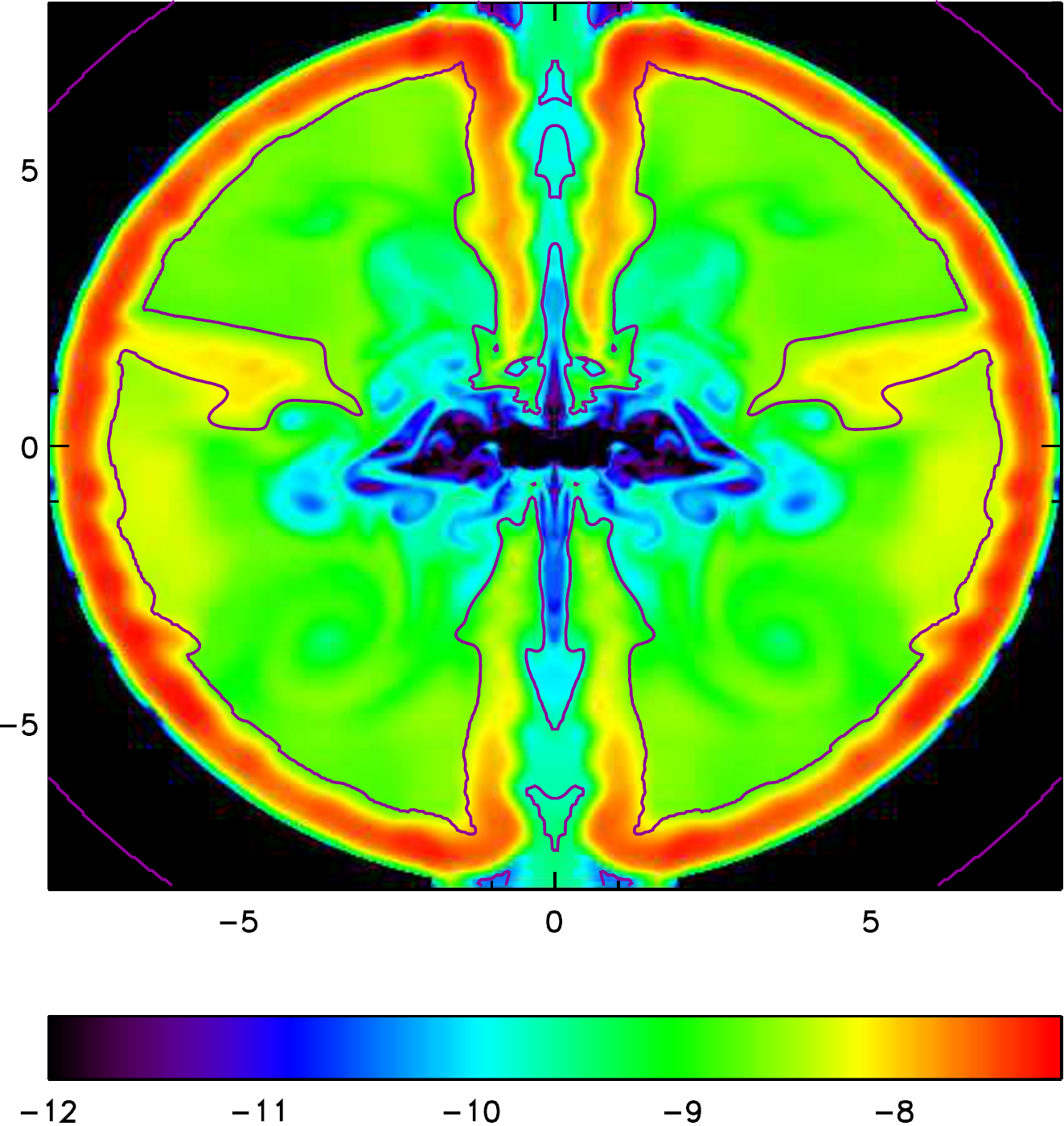}
 \hspace{1mm}
 \includegraphics[scale=.35]{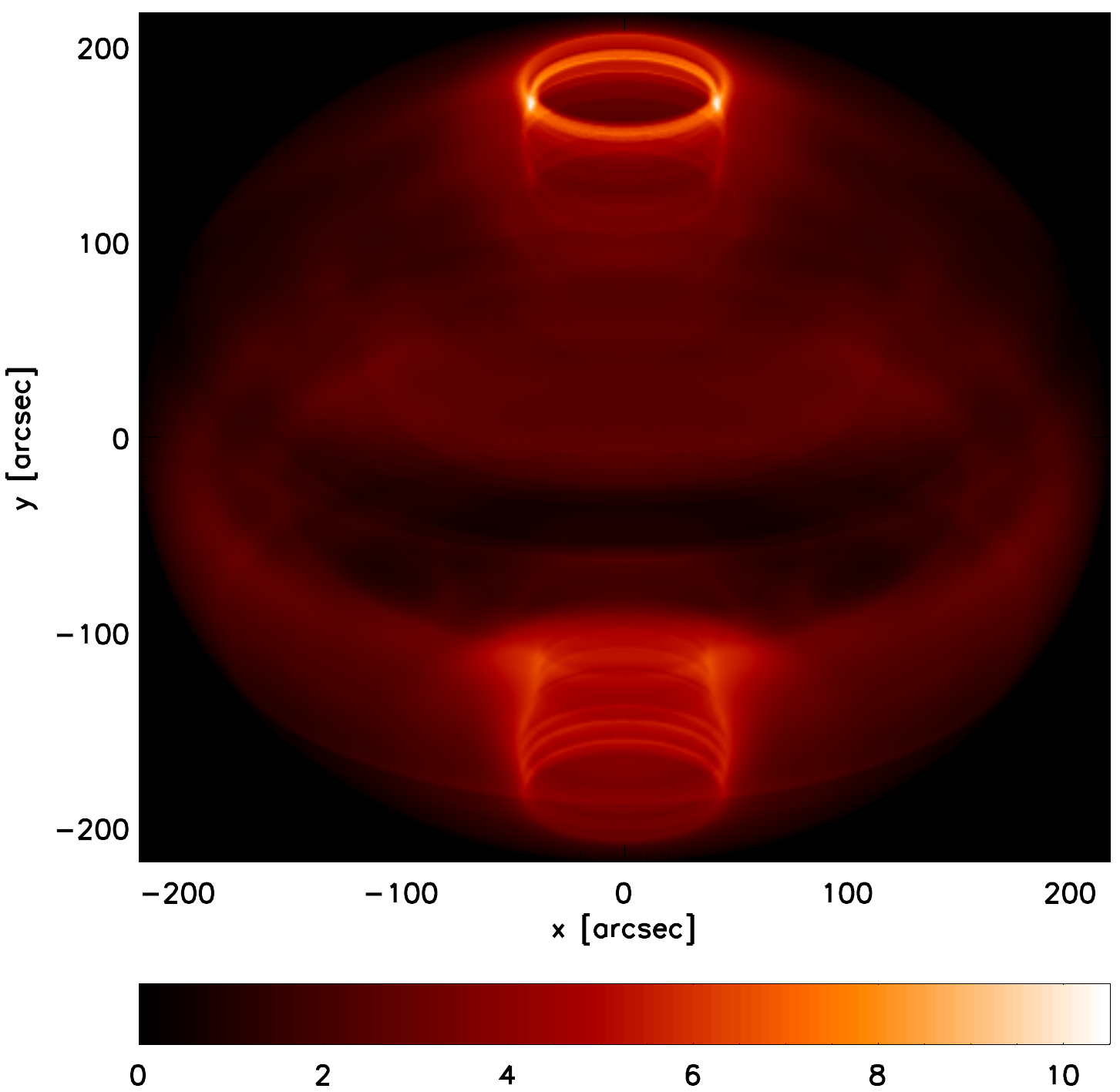}
 \caption{Radio particle density maps at $t=t_b=100$ yr (\textit{left panel}) and at $t=1000$ yr 
 (\textit{middle panel}); both maps are in logarithmic scale and on the $x$ and $y$ axes the 
 radial distance from the pulsar is reported in ly. The line contour in the second map traces the high 
 density region in the dynamic density map at the same stage of evolution, connected to the high density region in the tracer map: see text for more details.
  \textit{Right panel}: surface brightness map at 1.4 GHz, in linear scale and normalized 
 at the maximum value.}
 \label{fig:nr}
\end{figure*}

In Fig.~\ref{fig:nr} we show the density tracer at different times (left and middle panels) and the 
radio surface brightness map at $t=1000$~yr (right panel). In the image on the left, the density of radio 
particles is shown at the final stage of injection ($t=t_\mathrm{b}$), and its spatial distribution appears 
almost uniform in the nebula. 
The middle panel refers to the final stage of evolution, $t\simeq 1000$ yr, and two main differences 
are evident compared to the first map: the mean value is decreased by about three orders of magnitude 
and its distribution is not uniform anymore.The lower value of the density results from the adiabatic expansion of the nebular volume.

Furthermore, notice that a sort of higher density shell is visible at the outer edge of the nebula. 
This is caused by the accumulation of part of the dynamical mass density in that region in the early 
stages of evolution. Then vortices develop in the nebular flow pattern that confine the circulation in the body of the nebula, 
preventing new material from reaching the outer shell. As a consequence, the material that settled down there 
at the beginning does not get mixed anymore. 
The density of radio particles behaves exactly as mass density and this is the reason for the pattern that 
we observe in the figure. The same mechanism is at the base of the high density \emph{fingers} 
in the NW and NE regions of the nebula.

The line contour in the middle panel of Fig.~\ref{fig:nr} identifies the region of high mass density concentration. 
The radio surface brightness map in the right panel of Fig.~\ref{fig:nr} 
is computed with that region subtracted. In order to allow for direct comparison with the spatial distribution of particles, 
the emission map is not rotated in the plane of the sky, while is inclined with respect to the line of sight, 
so as to correctly take into account Doppler boosting.
Despite having removed most of the spurious effects of numerical origin, the emission morphology is clearly 
discrepant with respect to observations. The emission is diffuse, but the central region is much darker than observed. 
While the ring-like structures in the outer region of the nebula are likely related to residual spurious features near 
the impact regions of the jet, the absence of "wisp-like", Doppler boosted  structures in the inner region is a real 
difference between this map and the ones shown in the left panel of Fig.~\ref{fig:mapR}. This difference arises 
from the fact that there are no radio emitting particles left at the current time in the innermost region of the nebula, 
where the field strength is highest and the flow is most dynamic and with the fastest speed.

We have also tried different durations of the burst, while with shorter durations the emission spectrum 
does not extend to sufficiently high energies, a factor 2-3 longer duration leads to qualitatively similar results.
What we learn from this simple model, then, is that the radio particles cannot be described as a relic population 
of pulsar origin without including a local re-acceleration mechanism in the nebula.

\section{Comments and conclusion}
\label{sec:concl}

The axisymmetric MHD modeling of the dynamics of Pulsar Wind Nebulae is generally 
regarded as very successful. In this article we point out some of the limits it shows when 
its predictions are compared quantitatively with multi-wavelength observations of the Crab Nebula,
 the class representative that offers, thanks to the abundance of data, the tightest constraints.
The striking agreement between the simulated and observed high energy morphology of the 
Crab Nebula suggests indeed that the flow structure in the inner regions of this object must 
be very close to what our simulations predict \citep{Del-Zanna:2006}. On the other hand the same 
comparison suggests that the magnetic field cannot be perfectly toroidal all the way to the outer 
boundary of the nebula: some degree of turbulence is required in order to reproduce 
the brightness contrast between the \emph{Chandra} inner ring and torus. 
This could result from the development of kink instabilities that are artificially 
suppressed in our 2D simulations, and the same could be responsible 
for reducing the hoop stress that forces the wind magnetization estimated in 
2D to very low values. This very low magnetization is in fact incompatible with the 
integrated emission spectrum, as already highlighted by \citet{Volpi:2008}: reproducing the
synchrotron part of the spectrum requires conversion into accelerated particles 
with efficiencies larger than one, and the IC part cannot be reproduced at all.
This fact actually prevents us from drawing firm conclusions on the overall particle spectrum 
underlying the nebular emission. 

Recent works about current driven instabilities in 3D \citep{Mizuno:2011,Mignone:2013}
confirm the formation of kinks in jets, though the flow arising from the termination
shock and the hoop stresses in the nebula itself are not taken into account.
Preliminary work with full 3D relativistic MHD modeling of the entire system 
\citep{Porth:2013, Porth:2013a} seems to finally provide a solution to the long-standing $\sigma$ paradox:
since magnetic flux can be effectively destroyed in the body of the nebula, thanks to the development of 
kink instabilities, even an ultra-relativistic pulsar wind at equipartition allows for the formation of a PWN 
with features similar to those observed in the Crab Nebula, although the jets are now very weak and almost invisible. 
Whether all problems of PWN modeling are solved by moving from a 2D to a 3D relativistic MHD description 
is too early to assess. Current 3D simulations only extend for a very short time ($\sim 70$ yr) after the SN explosion, 
and at the end of the simulation the self-similar expansion phase has not been reached yet. Even more critical 
in our view is the fact that the average magnetic field in the nebula is already around $100 \mu G$ 
at this very young age, and likely to decrease by a factor larger than 10 by the time the simulated nebula 
gets to the Crab current age. No strong conclusions can be drawn at this stage.

The novelty of the present work resides in the first attempt ever at a detailed study 
of the low energy emission within the framework of 2D MHD. The main conclusion is that 
the radio emission morphology is basically insensitive to the spatial distribution of the particles 
in the two scenarios that see them either advected with the flow from the pulsar wind termination shock, 
or uniformly distributed in the nebula. Not only the surface brightness maps are basically identical, 
but also moving \emph{radio wisps} appear in both cases, and actually both the innermost fast moving 
features and the larger scale slower ones, in agreement with the observations by \cite{Bietenholz:2004}. 
In terms of variability, we compared the results of our simulations with the cited observations. 
The fact that the variable structures are well reproduced in our simulations independently of the spatial 
distribution of the particles points toward the conclusion that these details cannot really prove 
the origin of the radio emitting particles. 

The only scenario that can be excluded is one in which the low energy particles are  ``pure relics'', 
injected at early times after the SN explosion and then evolved in the absence of further acceleration 
and spatial diffusion. We were able to directly prove that this scenario does not work in the case of particle 
injection for less than 400 yr during the history of the nebula. In reality, the requirement on the last episode 
of acceleration of radio emitting particles would be even more stringent (closer in time episode) 
if the magnetic field in the simulated nebula were higher. In summary, radio emitting particles 
could only be fossil in terms of their injection in the nebula, but not in terms of their acceleration history. 

As far as acceleration mechanisms are concerned, apart from shock acceleration, the low energy particles could, 
in principle, be re-accelerated (if coming from the shock, or accelerated if coming from the thermal filaments) 
by distributed turbulence in the 
nebula, either of kinetic origin, injected by the higher energy population of particles, or of MHD origin.
Two main sources can be envisioned for the latter: the highly dynamical behavior of the shock front 
and nearby eddies where small-scale flows easily form and have been even invoked as sources 
of the gamma-ray flaring activity \citep{Komissarov:2011}, or the growth of local Kelvin-Helmoltz MHD 
instabilities at the shear layers of neighboring \emph{wisps}  \citep{Bucciantini:2006a}. 
Another possibility, actually supported by the results of \cite{Porth:2013} is that effective magnetic reconnection 
in the body of the nebula might also provide particle acceleration: this hypothesis is particularly attractive 
also because this mechanism naturally seems to provide flat energy spectra, just as generally inferred for young PWNe.

The possibility of explaining radio observations without requiring that the radio emitting particles 
must be part of the pulsar wind relaxes somewhat the strong
requirements on the multiplicity of the Crab pulsar found by \citet{Bucciantini:2011}. However,
explaining the very large number of these particles is still problematic and deserves further 
investigation. If they are of pulsar origin, then the pulsar multiplicity must have been high at
some point in time, and then have possibly decreased to accommodate for current observations
of X-ray emitting particles. On the other hand, the radio
emitters could also be produced elsewhere in the nebula, for example in the thermal filaments
due to Rayleigh-Taylor instabilities \citep{Bucciantini:2004a},
in which case they would only be electrons rather than pairs. In this case,
contrary to recent claims \citep{Blasi:2011}, PWNe could not be responsible 
for the \emph{positron excess} recently detected \citep{PAMELA-coll.:2009, AMS-02-coll.:2013}, 
\citep[for a review see also][]{Serpico:2012}. 
This explanation is not very appealing, especially because
very similar radio spectra are observed in bow-shock PWNe \citep[e.g.][]{Ng:2012}, and 
in these sources the particles are extremely unlikely to come from anywhere else than the pulsar. 

\footnotesize{
\bibliographystyle{mn2e}
\bibliography{olmi}
}

\label{lastpage}

\end{document}